\documentstyle[epsf,floats,eqsecnum,twocolumn,prb,aps]{revtex}
\newcommand{\s}{\mbox{\boldmath $S$}}
\newcommand{\bra}[1]{\langle#1|}
\newcommand{\ket}[1]{|#1\rangle}
\newcommand{\wg}[1]{\widetilde g_{#1}} 
\newcommand{\dg}[1]{\frac{dg_{#1}}{dl}}
\newcommand{\bN}{\mbox{\boldmath $N$}}
\newcommand{\bS}{\mbox{\boldmath $S$}}
\newcommand{\bn}{\mbox{\boldmath $n$}}
\newcommand{\bm}{\mbox{\boldmath $m$}}
\newcommand{\bl}{\mbox{\boldmath $l$}}
\begin{document}
\draft
\preprint{}
\title{Low-Energy Properties of Regularly Depleted Spin Ladders}
\author{ Takahiro Fukui\cite{Email}} 
\address{Institute of Advanced Energy, Kyoto University,
Uji, Kyoto 611, Japan}
\author{Manfred Sigrist \cite{leave}}
\address{ Yukawa Institute for Theoretical Physics, Kyoto University,
Kyoto 606-01, Japan  }
\author{Norio Kawakami}
\address{Department of Applied Physics,
%%%%and Department of Material and Life Science,\\
Osaka University, Suita, Osaka 565, Japan}
\date{October 31, 1996: Revised January 30, 1997; April 2, 1997}
\maketitle
%----------------------------------------------------------------------
%                               abstract
%----------------------------------------------------------------------
\begin{abstract}
We investigate a model for the regularly depleted two-leg 
spin ladder systems. By using Lieb-Schultz-Mattis theorem, 
it is rigorously shown that 
this model realizes massless excitations or, 
alternatively, a degenerate ground state,
although the original spin ladder system has a spin gap and
a unique ground state.
The ground state of the depleted model is 
either a spin singlet or partially ferromagnetic reflecting
topological properties of the depleted sites.
In order to show that the low-energy excitations are 
indeed massless, we 
proceed our analysis in two different ways
by resorting to effective field theories.
We first investigate an effective weak-coupling model
in terms of renormalization group methods.
%%%%%%%%%%%%%%%%%%%%%%%%%%%%%%%%%%%%%%%%%%%%%%%%%%%%%%%%
Although the tendency to massless spin excitations
is seen in the strong coupling regime, it turns out
that the model is still massive for any finite
coupling, implying
that a conventional weak-coupling approach is not efficient
to describe massless modes in our model. To overcome this difficulty,
%%%%%%%%%%%%%%%%%%%%%%%%%%%%%%%%%%%%%%%%%%%%%%%%%%%%% 
we further study low-energy properties of the
depleted spin model by mapping on the non-linear sigma model,
and confirm that the massless spin excitation 
indeed occur.
\end{abstract}

%----------------------------------------------------------------------
%                             introduction
%----------------------------------------------------------------------
\section{Introduction}

Since many decades one-dimensional quantum spin models have been a prominent
subject of theoretical study, representing a simple, but very rich class of
many-body systems. During the last three years the problem of several
coupled quantum spin chains has attracted much
attention.\cite{DAGOTTO} One reason 
lies certainly in the discovery of quasi one-dimensional spin systems in layered
cuprate materials which are the basis of high-temperature
superconductors. Synthesized under high pressure the
CuO$_2$-planes of the infinite layer compound SrCuO$_2$ creates 
regular line defects which lead to nearly decoupled one-dimensional spin
systems with ladder structure. There is a homologous
series of compounds, Sr$_{n-1}$ Cu$_{n+1}$ O$_{2n}$, which form
ladders with different numbers of legs (= (n+1)/2) depending on $ n
$.\cite{HIROI} The only (two-leg) spin ladder
compound previously known was (VO)$_2$ P$_2$ O$_7$. \cite{JOHNSTON} 

Spin 1/2 ladder systems with antiferromagnetic (AF)
nearest neighbor coupling
behave differently if they have an even or an
odd number of ladder legs. In case of an even number of legs the system has a
resonating valence bond (RVB) ground state (short-range singlet
correlation) and a gap to the lowest excitations. On the other hand,
an odd number of legs leads to gapless excitations and a ground state
with quasi long-range order similar to that of the single AF spin 1/2
chain.\cite{RICE} These properties have been clearly observed in experiments
for the compounds SrCu$_2$O$_3$ (two-leg ladder) and Sr$_2$Cu$_3$O$_5$
(three-leg ladder) and the size of the excitation gap for the two-leg
ladder system is in good agreement with theoretical predictions.\cite{AZUMA}

An interesting new aspect occurs if these systems contain non-magnetic
impurities, i.e., some spins are removed from the ladder. This can be
achieved, for example, by substituting Cu-ions by non-magnetic Zn.
Experiments showed that even a rather small concentration of Zn ($>1 \% $)
is sufficient to yield a transition to antiferromagnetic long-range
order in this two-leg ladder compound  SrCu$_2$O$_3$ as observed in recent
experiments. \cite{NOHARA} In model calculations it was actually
demonstrated that in  
the vicinity of a removed spin a staggered magnetization develops
which extends over several lattice constants. \cite{FUKUYAMA} Each
removed spin leaves locally an unpaired spin 1/2 degree of freedom behind.
These residual spins interact with each other through exchange of 
excitations of the ladder system analogous to the
RKKY-interaction. It can be shown easily that if two 
impurities lie on the same sublattice (ladders are bipartite lattices)
then the effective interaction between the corresponding residual spins is
ferromagnetic (FM). On the other hand the interaction is AF
if the two impurities occupy different sublattices. \cite{FURUSAKI} 
In the ground state these residual impurity spins correlate
due to their effective interaction. The orientation of the local
staggered moments associated with each impurity is also correlated with
that of the impurity spin. One can easily see that this 
leads to an in-phase alignment of the local staggered moments 
throughout the whole sample. In this context it is crucial that 
this system has no frustrating interactions. With
coupling among the ladders this behavior is indeed sufficient to create
AF long-range order. 
For the coherence of the AF correlation it does not matter whether the
impurities are located in a regular or random way along the
ladder. However, it is still not 
clear yet whether and under which conditions
true long-range order would also emerge in the
ground state of a single ladder. \cite{FURUSAKI,NAGAOSA} 

While the pure two-leg spin ladder has an excitation gap, 
a finite impurity concentration  seems to yield gapless excitations. This
was argued recently for the random depletion of spins, where for low
energies the system can be
reduced to an effective random spin chain.\cite{FURUSAKI} 
More recently, Iino et al. \cite{IINO} and Motome et al. \cite{MOTOME} 
analyzed the excitation spectrum
numerically. For very small impurity concentrations the spin
gap feature of the two-leg ladder system is still a dominant structure
and only few low-lying excitations appear far below the gap. However,
already at a concentration as small as $ 4 \% $ a crossover to a
regime occurs where the original excitation gap has essentially
disappeared.\cite{IINO,MOTOME} 

In this paper we would like to consider the problem of low-lying
excitations on a more rigorous basis for the case of the 
regular arrangement of impurities in two-leg ladder systems. 
In sec II, we first
 prove rigorously by the Lieb-Schultz-Mattis theorem 
that any finite concentration of removed spins leads to
an excited state with the energy of $O(1/N)$, 
which implies the formation of either gapless excitations or 
the existence of a degenerate ground state with  spin gap
($N$: number of lattice sites of the finite system). 
In the next step, we will attempt to determine which possibility
is actually realized in the model by  
field-theoretical methods. For this purpose we first introduce 
in sec.III an effective Hubbard-type ladder model  
and study its low-energy properties using the 
renormalization group method. In this Hubbard model the effective
depletion of spins is incorporated in the strong coupling limit of 
on-site interaction which 
is introduced to produce the singlet bound states at 
each depleted lattice point. 
%%%%%%%%%%%%%%%%%%%%%%%%%%%%%%%%%%%%%%%%%%
Though the tendency to gapless excitations 
is seen as the singlet bound state becomes strong, it turns out
that the model is still massive for any finite couplings.
This may suggest  that it is not straightforward to
incorporate the effects of depletion 
in a conventional continuum limit.
%%%%%%%%%%%%%%%%%%%%%%%%%%%%%%%%%%%
To clarify this point, we further study in sec. IV 
the depleted spin model by employing  a complementary field
theoretical approach, i.e. the non-linear sigma model.
%%Our analysis here may be valid for the low concentration of depletion.
To this end, we first calculate the dispersion 
of the spin waves by the Holstein-Primakoff mapping. We then
find that in the case of singlet (partially ferromagnetic) 
ground state, the spectrum of the lowest band is  linear (quadratic)
without a gap, suggesting that correlations between unpaired 
spins are in fact essential for  low-energy excitations.
Based on this observation, we study 
the model with singlet ground state by 
%%a complementary approach which may be valid 
%%for the high concentration of depletion, i.e,
using the mapping to the non-linear sigma model.
%%for the purpose of studying the quantum effects on the massless free
%%spin wave excitation.
It is found that the coefficient of the topological term 
in the effective sigma model is 
$\pi i$, analogous to the spin 1/2 antiferromagnetic Heisenberg chain, 
suggesting that our spin ladder systems with 
periodic depletion indeed have massless spin excitations,
being consistent with the analyses in sec.II and III.
Brief summary is given in sec. V.

We wish to mention here 
that the model studied numerically by Iino and Imada\cite{IINO} is
similar to the one we introduce in Sec.II. They discussed
how the massless states appear as the impurity
concentration is increased. Our approach based on the 
rigorous statements
may provide the results complementary to theirs, and also via
the present analysis 
we will clearly see how the coherence of the eigenfunction 
is modified by the depletion.
%%%%%%%%%%%%%%%%%%%%%%%%%%%%%%%%%%%%%%%%%%%%%%%%%%%%%%%

%-----------------------------------------------------------------------
%                            II.A
%-----------------------------------------------------------------------
\section{Regularly Depleted Spin Ladder}
\subsection{Model}

\begin{figure}[h]
\epsfxsize=7cm %%% 8 is suitable for 2 column
\centerline{\epsfbox{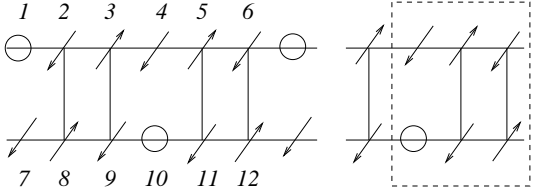}} 
\vspace{0.5cm}
\caption{Regularly-depleted two-leg spin ladder system with $m=6$.
The number assigned to each site is used 
in the spin-wave analysis in section IV.A.}
\label{fig:SpiLad}%-----------------------------------------------------
\end{figure}

As mentioned above, the presence of the gapless state 
is a quite common feature in the 
depleted spin 
ladders.\cite{RICE,AZUMA,NOHARA,FUKUYAMA,FURUSAKI,NAGAOSA,IINO,MOTOME}
Although this has been already claimed by various 
studies based on impurity models,
it may be important to rigorously prove
that such a gapless state can be indeed realized by the depletion.
In order to study the effects of depleted spins on 
two-leg ladder systems, we study here a special 
class of the spin models, i.e.,  regularly depleted
spin ladder systems, from which one can clearly see
the drastic change of the ground state.
The model Hamiltonian we consider is
%-----------------------------------------------------------------------
%   hamiltonian 
%-----------------------------------------------------------------------
\begin{eqnarray}
H=&&H^{(1)}+H^{(2)}+H_{\rm coup},\nonumber\\
&&H^{(i)}=J_\|\sum_{j=0}^{N-1}
(1-\theta_j^{(i)})(1-\theta_{j+1}^{(i)})
\s_j^{(i)}\cdot\s_{j+1}^{(i)},\nonumber\\
&&H_{\rm coup}=J_\bot\sum_{j=0}^{N-1}
(1-\theta_j^{(1)})(1-\theta_{j}^{(2)})
\s_j^{(1)}\cdot\s_{j}^{(2)},
\label{Ham}%------------------------------------------------------------
\end{eqnarray}
where $\theta$ is defined by $\theta_j^{(1)}=\theta_j$
and $\theta_j^{(2)}=\theta_{j+m/2}$ with
%-----------------------------------------------------------------------
%   theta
%-----------------------------------------------------------------------
\begin{equation}
\theta_j=\left\{
\begin{array}{ll}1&\hbox{ for }j=0 \hbox{ (mod }m)\\
0&\hbox{ otherwise }\end{array}\right..
\label{TheJ}%--------------------------------------------------------
\end{equation}
In what follows, we set $J_\|=1$ and $J_\bot=J$ for simplicity.
This model is defined on the two coupled chains, 
each of which consists of 
$N$ sites labeled by $j=0,1,\cdots,N-1$. An
even-integer $m$ denotes the period of depleted sites
(we call them impurities) with $N=mM$, 
where $M$ denotes the number of impurities on each chain.
Periodic boundary conditions are imposed on the system.
We show the model in Fig. \ref{fig:SpiLad}
schematically for the case of $m=6$.
%----------------------------------------------------------------------
%   Fig.1
%-----------------------------------------------------------------------

Although our model seems rather special,
it is to be noted that the model contains  an essential property 
expected generally for the depleted models\cite{IINO}: 
the gapless states are essentially formed by 
the coherent motion of unpaired spins
generated  by the depletion,\cite{FUKUYAMA} 
though an unpaired spin may be 
a rather complicated object in general.
For example, in the case where the period of the depletion is 
large (dilute limit of impurities), our model reproduces the known
results for the system with a few non-magnetic impurities.  
But also for the high density limit, 
we can discuss what is essential for the formation of a 
gapless state in the depleted systems. 
Moreover, since we can give a rigorous statement 
for the model, it may serve as a key model by which we can check 
whether an approximate treatment works well or not when we 
study wider class of depleted ladder systems.
In fact, we will treat  a slightly different model 
in the next section, for which the exact results obtained 
in this section provide a guideline for 
analyzing the results correctly.

%-------------------------------------------------------------------
%                          II.B 
%------------------------------------------------------------------
\subsection{The Lieb, Schultz and Mattis Theorem}

By using the Lieb, Schultz and Mattis (LSM) theorem
\cite{LSM},  we now prove that
the depleted model (\ref{Ham}) indeed has an excited state of $O(1/N)$.
The LSM theorem is elementary, 
but provides quite powerful tool to demonstrate that  
a given system is gapless or has degenerate ground states.
This theorem can be applied 
not only to half-odd integer spin models\cite{LSM,AffLie} but also 
to the  spin ladder systems, which suggests 
that odd-leg ladder models are expected to have massless excitations,
while it is not applicable to even-leg ladder models,
indicating the presences of an excitation gap. 
The relation between the LSM scheme and the massive RVB states was
discussed previously in detail in literature (see for example
Ref.14 and references therein).

We now wish to apply this theorem to the 
depleted model (\ref{Ham}). The theorem uses two properties. 
One is the symmetry of the model based on the translation combined
with reflection.
The other is the response of the ground state energy 
to twisted boundary conditions, or equivalently to external
gauge fields.  This plays an important role for studying 
low-energy properties in quantum many-body systems. 
In particular, such effects appear in quite different ways for massless
and massive systems.  For example, for massless systems,
 the ground state energy  
increases as a function of the twist angle of $O(1/N)$, 
where $N$ is the size of the
system, so that  the ground state changes 
into one of the excited states when we follow the spectral flow of the
ground state up to the twist angle $\Phi=2\pi$\cite{ABB,SutSha}.
On the other hand, for massive systems, there is a gap 
above the ground state of 
$O(1)$, and hence we cannot reach the excited states by twisting the
ground state.  In this way, we can use twisted 
boundary conditions to find a gapless excitation.

We now turn to the question whether
we can find such an excitation of $O(1/N)$
in the present model. 
To this end, let us introduce the 
operator $T_{m/2}$ for $m/2$-lattice 
translation and the reflection operator $R$, 
%----------------------------------------------------------------------
%   translation and reflection
%-----------------------------------------------------------------------
\begin{eqnarray}
&&T_{m/2}\s_j^{(i)}T_{m/2}^{-1}=\s_{j+m/2}^{(i)},\nonumber\\
&&R\s_j^{(1,2)}R^{-1}=\s_j^{(2,1)}.
\end{eqnarray}
Let $O$ be the product operator $O=RT_{m/2}$.
The total Hamiltonian is invariant under
the operation of $O$ since $OH^{(1,2)}O^{-1}=H^{(2,1)}$ and 
$OH_{\rm coup}O^{-1}=H_{\rm coup}$.
On the other hand, for a finite size system
the ground state is non degenerate within the
subspace of fixed $ S^z_{tot} $ ($=0$), so that we 
can write $O\ket{\Psi_0}=e^{i\alpha}\ket{\Psi_0}$.

Now we introduce the twist operator, which plays the central role in
the LSM theorem,
%------------------------------------------------------------------------
%   twist operator
%-----------------------------------------------------------------------
\begin{equation}
U^{(i)}=\exp\left[\frac{2\pi i}{N}\sum_{j=0}^{N-1}(1-\theta_j^{(i)})
jS_j^{(i)z}\right].
\end{equation}
This operator transforms under $O$ such that
%-----------------------------------------------------------------------
%   transformation of each twist op.
%----------------------------------------------------------------------
\begin{eqnarray}
OU^{(1)}O^{-1}=&&U^{(2)}\exp\left(-\frac{m\pi i}{N}S^{(2)z}\right)
\nonumber\\
&&\times\exp\left[2\pi i\sum_{j=0}^{m/2-1}
(1-\theta_j^{(2)})S_j^{(2)z}\right],
\nonumber\\
OU^{(2)}O^{-1}=&&U^{(1)}\exp\left(-\frac{m\pi i}{N}S^{(1)z}\right),
\nonumber\\
&&\times\exp\left[2\pi
i\sum_{j=0}^{m/2-1}(1-\theta_j^{(1)})S_j^{(1)z}\right] ,
\end{eqnarray}
and therefore $U=U^{(1)}U^{(2)}$ is transformed as
%------------------------------------------------------------------------
%   transformation of total twist op
%-----------------------------------------------------------------------
\begin{eqnarray}
OUO^{-1}=&&U\exp\left(-\frac{m\pi i}{N}S^{z}\right)
\nonumber\\
&&\times\exp\left[2\pi i\sum_{i=1}^2\sum_{j=0}^{m/2-1}
(1-\theta_j^{(i)})S_j^{(i)z}\right]\nonumber\\
=&&-U .
\end{eqnarray}
The second line comes from the facts that 
we are concerned with   the  total spin  $S^z=0$ sector and 
that  in the last 
exponential there are $m-1$ (odd) spins
inside of the dashed box in Fig.\ref{fig:SpiLad}, each of
which gives the factor $-1$.
Owing to this property, we can construct an excited 
state $\ket{\Psi}=U\ket{\Psi_0}$ 
with different ``parity'', which is orthogonal to the ground state,
$\bra{\Psi_0}\Psi\rangle=0$.

The remaining task is to calculate the energy increment $\delta
E=\bra{\Psi}H\ket{\Psi} -\bra{\Psi_0}H\ket{\Psi_0}$.
We have
%-----------------------------------------------------------------------
%   energy increment
%-----------------------------------------------------------------------
\begin{eqnarray}
\delta E=&&\sum_{i=1}^{2}\sum_{j=0}^{N-1}
(1-\theta_j^{(i)})(1-\theta_{j+1}^{(i)})\nonumber\\
&&\times\left(\cos\frac{2\pi}{N}-1\right)
\bra{\Psi_0}S_j^{(i)+}S_j^{(i)-}\ket{\Psi_0}\nonumber\\
&&\le 2(N-M)\left(1-\cos\frac{2\pi}{N}\right)
<2\left(1-\frac{2}{m}\right)\frac{4\pi^2}{N}.
\end{eqnarray}
Namely, the excited state orthogonal to the 
ground state has the excitation energy of $O(1/N)$.
Therefore, we end up with the rigorous statement 
that our model (\ref{Ham}) for regularly depleted 
ladder systems has either the massless
spin excitations or the ground state is degenerate in 
the thermodynamic limit.  
In subsequent sections, we will focus our attention to the question 
whether low-energy excitations of our system are indeed massless,
and, if so, which type of spin excitations contributes to this 
massless mode.
 
Before concluding this section, we wish to remark the
following points.
As mentioned above, it is the most crucial point in the LSM theorem
whether we can construct a state orthogonal to the ground 
state by the use of the twist operator.
In the present system, an unpaired spin in the dashed box in
Fig.\ref{fig:SpiLad} gives the factor $-1$.
In the ordinary two-leg ladder systems without depletion, we have always 
an even number of spins inside the dashed box (however, in this case
translation by one site is sufficient to consider, which still gives
the factor $(-1)^2=1$). Therefore, we can explicitly see
that a drastic change in the ground state directly 
reflects how  the phase coherence of the wave function
is changed by the depletion.

%----------------------------------------------------------------------
%   discussion of the spin of the ground state
%----------------------------------------------------------------------
Some aspects of the ground state can be discussed by applying
Marshall's theorem to our model.\cite{Aue,LM}
The ground state has the spin quantum number
$S=0~(M)$ for $m=4~(2)$ mod 4, so that 
the ground state of our model is 
either a spin singlet or partially ferromagnetic.
%%%%%%%%%%%%%%%%%%%%%%%%%%%%%%%%%%%%%%%%%%%%%%%%%%%%%%%
Therefore, in the partially ferromagnetic case, the ground state
itself is degenerate, since the state under consideration ($S^z=0$
state) is, of course, a member of the spin $S=M$ multiplet.
%%%%%%%%%%%%%%%%%%%%%%%%%%%%%%%%%%%%%%%%%%%%%%%%%  
As mentioned in the introduction, this can be easily understood
by noticing that the effective interaction
between the unpaired spins is ferromagnetic (antiferromagnetic)
if the depleted sites are on the same (different)
sublattice(s).\cite{FURUSAKI}  
%%In both cases, the massive state of the 
%%undepleted model is changed into the 
%%new gapless states.

%------------------------------------------------------------------------
%          III. Weak-coupling approach to Hubbard ladder
%-----------------------------------------------------------------------
\section{Weak-coupling approach to Hubbard ladder}

In the previous section, we have proved that our model for
the regularly depleted ladder systems is characterized by massless
spin excitations or degenerate ground states with a gap. 
In order to determine which possibility is actually realized
and to study, if possible, the nature of elementary 
excitations in more detail, we wish to introduce a low-energy effective 
theory in the continuum limit and study its low-energy 
properties by using the renormalization group (RG) method.
One immediately notices, however,  that 
it is not straightforward to study the effect of the 
periodic depletion
in an ordinary bosonization approach. 
For example, we cannot take a conventional way in which
each chain is first bosonized and the inter-chain coupling is then 
taken into account  via the RG procedure, because a bare 
single chain would be divided 
into many disconnected pieces by the  depleted sites. 
To avoid this difficulty, it is desirable to find a way 
to include the effect of the periodic depletion after bosonization.
Based on these observations we introduce a 
Hubbard-type ladder model in which 
the effect of  the depletion is incorporated in terms of the 
on-site interaction.  We then investigate its  
low-energy properties by the one-loop RG method.
We will use the notations similar to those of Balants and
Fisher\cite{Fisher}, who studied the ordinary Hubbard type ladder
systems without depletion.

%----------------------------------------------------------------------
%                   III.A
%----------------------------------------------------------------------
\subsection{Model}

\begin{figure}[h]
\epsfxsize=5cm %%% 8 is suitable for 2 column
\centerline{\epsfbox{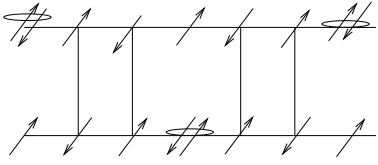}} 
\vspace{0.5cm}
\caption{Schematic illustration of the two-leg Hubbard ladder in the
strong-coupling limit $U'\rightarrow-\infty$}
\label{fg:HubLad}%------------------------------------------------------
\end{figure}

The Hamiltonian we consider is the two-leg ladder of
correlated electrons, 
%-----------------------------------------------------------------------
%   Hubbard ladder hamiltonian
%-----------------------------------------------------------------------
\begin{eqnarray}
H=&&\sum_{i=1}^2(H_{\rm hop}^{(i)}+H_{\rm int}^{(i)})+H_{\rm coup},
\nonumber\\
&&H_{\rm hop}^{(i)}=-t\sum_{j=0}^{N-1}\sum_\sigma
\left(c_{j\sigma}^{(i)\dagger}c_{j+1\sigma}^{(i)}+h.c.\right),
\nonumber\\
&&H_{\rm int}^{(i)}=\sum_{j=0}^{N-1}(U+U'\theta_j^{(i)})
n_{j\uparrow}^{(i)}n_{j\downarrow}^{(i)},
\nonumber\\
&&H_{\rm coup}=J\sum_{j=0}^{N-1}
(1-\theta_j^{(1)})(1-\theta_j^{(2)})\s_j^{(1)}\cdot\s_j^{(2)},
\label{HubHam}%-----------------------------------------------------
\end{eqnarray}
where we have introduced two types of the 
Hubbard interactions $U>0$ and $U+U'<0$.
The interaction $U$ is introduced to produce 
Heisenberg spins on the lattice sites for large  $U$. 
On the other hand, the attractive interaction $U'$ is introduced
to effectively represent the depleted spins.  Namely, 
in the $U+U'\rightarrow-\infty$ limit, 
electrons on $j=0~(m/2)$ mod $m$ sites for chain-1~(-2)
form bound states, i.e. onsite spin singlets,
 and hence the spin degrees of freedom are effectively 
frozen out on these sites, as illustrated in Fig.\ref{fg:HubLad}.  
In order to correctly reproduce the depleted ladder systems,
the number of electrons should be $2(N+M)$,
and then the corresponding Fermi momentum is 
%---------------------------------------------------------------------
%   Fermi momentum
%---------------------------------------------------------------------
\begin{equation}
k_F=\frac{\pi}{2a_0}\left(1+\frac{1}{m}\right),
\end{equation}
where $a_0$ is the lattice constant.

%------------------------------------------------------------------------
%   Fig. 2
%------------------------------------------------------------------------

We now investigate low-energy properties 
of this model with the RG method.
Passing to the continuum limit, the fermion operators become 
%----------------------------------------------------------------------
%   continuum limit of the fermi op
%----------------------------------------------------------------------
\begin{equation}
c_{j\sigma}^{(i)}/\sqrt{a_0}\rightarrow 
e^{-ik_Fx}\psi_{iL\sigma}(x)+e^{ik_Fx}\psi_{iR\sigma}(x)
\end{equation}
with $aj\rightarrow x$,
and the hopping Hamiltonian is reduced to
%------------------------------------------------------------------------
%   continuum limit of the free hamiltonian 
%-----------------------------------------------------------------------
\begin{eqnarray}
H_{\rm hop}^{(i)}=v_F\int dx
\Bigg[&&\frac{\pi}{8}(J_{iL}^0J_{iL}^0+J_{iR}^0J_{iR}^0)
\nonumber\\
&&+\frac{2\pi}{3}(J_{iL}^aJ_{iL}^a+J_{iR}^aJ_{iR}^a)
\Bigg]
\end{eqnarray}
where $v_F=2ta_0\cos(\pi/2m)$. 
Here and in what follows, operator products should be normal-ordered,
though we do not explicitly indicate this. 
$J_{iL(R)}^\mu$ is the left (right)
component of the U(1) and SU(2) currents defined by
%------------------------------------------------------------------------
%   definition of currents
%-------------------------------------------------------------------------
\begin{equation}
J_{iL(R)}^\mu(x)=\psi_{iL(R)\alpha}^\dagger
\frac{\sigma^\mu_{\alpha\beta}}{2}
\psi_{iL(R)\beta}(x),
\end{equation}
where $\sigma^0$ is the unit matrix: The component $\mu=0$ 
is the usual U(1) current divided by 2, 
and those with $a=\mu=1,2,3$ are the SU(2)
currents, respectively. Next we bosonize 
the Hubbard interaction in this approximation.
In the following we neglect the oscillating terms which are
incommensurate with the Fermi momentum and $\theta_j^{(i)}$, as
they may disappear after the integration. 
According to calculations briefly summarized in Appendix A,
the intra-chain interaction leads to
the following Hamiltonian density,
%------------------------------------------------------------------------
%   Hubbard interaction
%-----------------------------------------------------------------------
\begin{eqnarray}
-{\cal H}_{\rm int}=&&
 \wg{\rho}J_{iL}^0J_{iR}^0
+\wg{x\rho}(J_{1L}^0J_{2R}^0+J_{2L}^0J_{1R}^0)\nonumber\\
&&+\wg{u}(-)^{i-1}(M_{iL}M_{iR}^\dagger+M_{iL}^\dagger
M_{iR})\nonumber\\
&&+\wg{\sigma}J_{iL}^aJ_{iR}^a ,
\label{HubInt}%-------------------------------------------------------
\end{eqnarray}
where
%----------------------------------------------------------------------
%   definition of M_{iL}
%-------------------------------------------------------------------
\begin{equation}
M_{iL(R)}(x)=\psi_{iL(R)\downarrow}\psi_{iL(R)\uparrow},
\end{equation}
and we have dropped the chiral interactions because they 
merely  renormalize the Fermi velocities. 
Here, repeated indices $i$ and $a$ are summed over.
Note that the Umklapp interaction appears though 
the density of the system 
under consideration is more than half-filling.
This is due to the existence of $\theta_j$ in eq.(\ref{HubHam}), 
which fixes $U'$ on regularly placed lattice sites.
We can also take the continuum 
limit of the inter-chain coupling
%-----------------------------------------------------------------------
%   inter-chain coupling
%-----------------------------------------------------------------------
\begin{eqnarray}
-{\cal H}_{\rm coup}=&&
\wg{x\sigma}(J_{1L}^aJ_{2R}^a+J_{2L}^aJ_{1L}^a)
+\wg{t\sigma}(L_L^aL_R^{a\dagger}+L_L^{a\dagger}L_R^a)\nonumber\\
&&+\wg{t\rho}(L_L^0L_R^{0\dagger}+L_L^{0\dagger}L_R^0)
\label{InterCha}%--------------------------------------------------
\end{eqnarray}
where $L_{L(R)}$ is the staggered components of the spin operator
%----------------------------------------------------------------------
%   definition of L op
%----------------------------------------------------------------------
\begin{equation}
L_{L(R)}^\mu=
\psi_{1L(R)\alpha}^\dagger
\frac{\sigma^\mu_{\alpha\beta}}{2}
\psi_{2L(R)\beta}
\end{equation}
%-----------------------------------------------------------------------
%   initial coupling
%-----------------------------------------------------------------------
The initial values of these
coupling constants are given by the Hubbard and inter-chain
interactions, as summarized in Table \ref{t:I},
where we have defined  $\widetilde U'=U'/m, 
\widetilde U=U+\widetilde U'$ and $\widetilde J=J(1-2/m)$.
%------------------------------------------------------------------------
%   Table I
%------------------------------------------------------------------------
\begin{table}[h] % 'e' is the last page.
\begin{center}
\renewcommand{\arraystretch}{2}
\begin{tabular}{ccccccc}
%\hline
$\wg{\rho}$&$\wg{x\rho}$&$\wg{u}$&$\wg{\sigma}$
&$\wg{x\sigma}$&$\wg{t\sigma}$&$\wg{t\rho}$\\
\hline%\multicolumn{7}{c}{}\\
$-2\widetilde Ua_0$ & 0 & $-\widetilde U'a_0$ & $2\widetilde Ua_0$ 
&$-\widetilde Ja_0$ & $-\frac{1}{2}\widetilde Ja_0$ 
& $\frac{3}{2}\widetilde Ja_0$\\
%\hline
\end{tabular}
\end{center}
\caption{Coupling constants}
\label{t:I}%------------------------------------------------------
\end{table}
The bosonized forms of the interactions are 
also listed in Appendix C.

%-----------------------------------------------------------------------
%                     III. B RG equation
%---------------------------------------------------------------------
\subsection{RG Equations and Flows}

We now derive the RG equations for the above coupling constants
in one-loop order.
Using the operator product expansions and resultant RG equations
given in Appendix B, 
we end up with the following set of scaling  equations,
%---------------------------------------------------------------------
%   RG equation
%--------------------------------------------------------------------
\begin{eqnarray}
&&\dg{\rho}=-4g_{u}^2-\frac{1}{4}(g_{t\rho}^2+3g_{t\sigma}^2),
\label{Grho}\\%--------------------------------------------------
&&\dg{x\rho}=\frac{1}{4}(g_{t\rho}^2+3g_{t\sigma}^2),
\label{GXrho}\\%-------------------------------------------------
&&\dg{u}=-g_\rho g_u,
\label{Gu}\\%-----------------------------------------------------
&&\dg{\sigma}=-g_\sigma^2
 -\frac{1}{2}(g_{t\sigma}+g_{t\rho})g_{t\sigma},
\label{Gsigma}\\%------------------------------------------------
&&\dg{x\sigma}=-g_{x\sigma}^2
 -\frac{1}{2}(g_{t\sigma}-g_{t\rho})g_{t\sigma},
\label{GXsigma}\\%-----------------------------------------------
&&\dg{t\sigma}=
-\left[\frac{1}{2}(g_\rho-g_{x\rho})
 +(g_\sigma+g_{x\sigma})\right]g_{t\sigma}\nonumber\\
 && \qquad\qquad\qquad
  -\frac{1}{2}(g_\sigma-g_{x\sigma})g_{t\rho},
\label{GTsigma}\\%----------------------------------------------
&&\dg{t\rho}=-\frac{1}{2}(g_\rho-g_{x\rho})g_{t\rho}
 -\frac{3}{2}(g_\sigma-g_{x\sigma})g_{t\sigma}
\label{GTrho}%----------------------------------------------------
\end{eqnarray}
with $l=\ln L$, where $g_k=\widetilde g_k/2\pi v_F$.

\begin{figure}[h]
\epsfxsize=6cm %%% 8 is suitable for 2 column
\centerline{\epsfbox{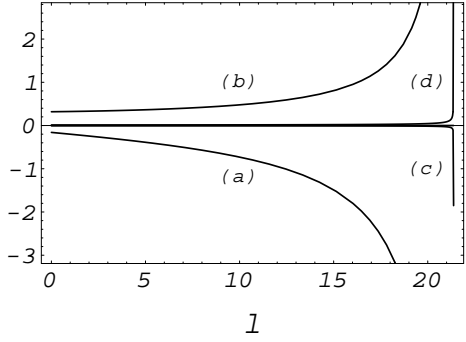}} 
\vspace{0.5cm}
\caption{RG flows for (a) $g_\rho$, (b) $g_{u}$, (c) $g_{t\sigma}$
and (d) $g_{t\rho}$: Plotted values are multiplied by the factor 10. 
$g_{x\rho}\sim0$ in this figure.
The parameters used are
$U=1, U'=-20, j=0.01$ and $m=50$.}
\label{fg:flow1}%----------------------------------------------------
\end{figure}

\begin{figure}[h]
\epsfxsize=6cm %%% 8 is suitable for 2 column
\centerline{\epsfbox{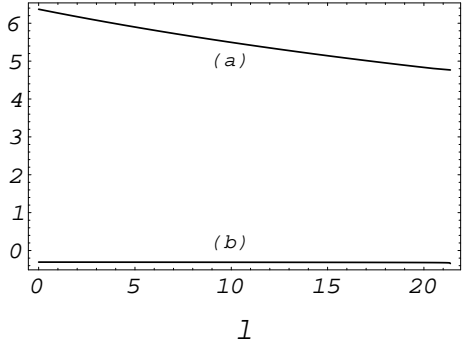}} 
\vspace{0.5cm}
\caption{RG flows for (a) $g_\sigma$ and (b) $g_{x\sigma}$:
Plotted values are multiplied by the
factor 400. The parameters used are the 
same as those in Fig.\ref{fg:flow1}.}
\label{fg:flow2}%----------------------------------------------------
\end{figure}

We have numerically integrated the above set of equations
to obtain the RG flows.
Typical examples of the RG flows are shown in Figs.\ref{fg:flow1} and
\ref{fg:flow2}.
%------------------------------------------------------------------------
%   Fig. 3,4
%------------------------------------------------------------------------
We note that a characteristic scale $l_c$ exists,
at which all renormalized couplings except for $g_{x\rho}$ exhibit
divergence properties, though those in Fig.\ref{fg:flow2} look
convergent at a first glance. 
We will see below that this divergence is mainly driven by
the charge sector, so that $l_c$ may be identified 
with the scale at which the mass gap is generated 
for the charge sector. Accordingly, couplings for other spin modes
are strongly affected. In what follows, we analyze  characteristic 
properties of RG flows around $l_c$ analytically.
Note first that  the key parameter 
in our approach is $U'$, which would make singlet bound pairs
on particular sites, freezing out the spin degrees of freedom.
Now we fix the other parameters, e.g. 
$U=1, m=50, J=0.01$. In this case, the Um\-klapp interaction
turns out to be relevant in the range $U'_c<U'<0$ (
$U'_c\sim -40$ for this particular choice of interactions.
We will explain the meaning of $U_c'$ below).
In this range, $g_{\rho}$ is most relevant and 
we can neglect $g_{t\sigma}$ and $g_{t\rho}$ 
eqs.(\ref{Grho})$\sim$(\ref{Gu}), resulting in 
%----------------------------------------------------------------
%   app. eq. for charge
%--------------------------------------------------------------
\begin{eqnarray}
\dg{\rho}&\sim&-4g_u^2,
\nonumber\\
\dg{u}&\sim&-g_\rho g_u ,
\end{eqnarray}
near the divergent point.
This is a familiar U(1) scaling equation, from which we obtain
$g_\rho\sim -1/(l_c-l)$. 
This expression near the divergent point is confirmed by the numerical 
integration of the RG equations.
This result means that the charge gap
is generated by the Coulomb interactions $U$ and $U'$.

\begin{figure}[h]
\epsfxsize=6cm %%% 8 is suitable for 2 column
\centerline{\epsfbox{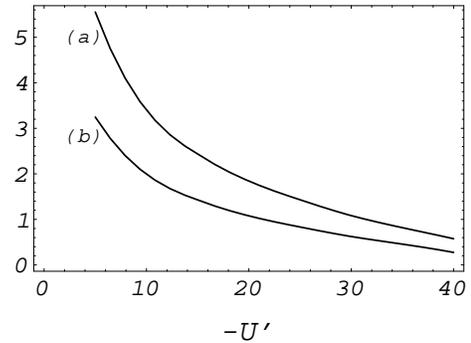}}
\vspace{0.5cm}
\caption{(a) $-g_{t\sigma0}\times 10^3$ and (b) 
$g_{t\rho0}\times2\cdot10^2$ as
functions of negative $U'$.}
\label{fig:Gt}%------------------------------------------------------
\end{figure}

What happens for the spin sector in this case ?
Note that the 
next relevant interactions are the inter-chain staggered interactions
$g_{t\sigma}$ and $g_{t\rho}$ for the spin sector,  since
they act as composite of spin and charge degrees of freedom, 
the dimension of which is  unity if we replace 
the massive charge degree of
freedom by its mean value, as can be seen in Table.\ref{t:Bos}.
Neglecting small terms, we thus obtain
%------------------------------------------------------------------
%   app. eq. for staggered spin
%----------------------------------------------------------------
\begin{eqnarray}
\dg{t\sigma}&\sim&-\frac{1}{2}g_\rho g_{t\sigma},
\nonumber\\
\dg{t\rho}&\sim&-\frac{1}{2}g_\rho g_{t\rho} .
\end{eqnarray}
Using the above-obtained $g_\rho$ for the 
charge sector, we have the RG flows 
%------------------------------------------------------------------
%   asymp. for staggered spin
%------------------------------------------------------------------
\begin{eqnarray}
g_{t\sigma}&\sim&\frac{g_{t\sigma0}}{(l_c-l)^{1/2}},
\nonumber\\
g_{t\rho}&\sim&\frac{g_{t\rho0}}{(l_c-l)^{1/2}},
\label{GTsol1}%-----------------------------------------------------
\end{eqnarray}
near the divergent point. Therefore, it is seen that 
the spin modes scale to the strong coupling fixed-point
 and the spin gap is still open. 
This may not be plausible for the depleted spin ladders,
according to the rigorous theorem
obtained in the previous section. This is because
the effect of the depletion is not fully taken into account in the 
present one-loop RG method.\cite{Com1}  
It should be noted, however, that 
the effect of the depletion actually has the tendency to
make the spin sector massless. Namely, we can indeed see
that the coefficients $g_{t\sigma0}$  
and $g_{t\rho0}$ quickly decrease with increasing 
$|U'|$. \cite{Com2}
In fact from the numerical integration of the RG equations, 
we estimate them as
%-------------------------------------------------------------------
%   coef. as fun. of U'
%------------------------------------------------------------------
\begin{equation}
g_{t\sigma0}, g_{t\rho0}\sim \frac{\mbox{const.}}{|U'|^\alpha},
\label{GT0}%---------------------------------------------------------
\end{equation}
with $\alpha\sim 1$ (See Fig.\ref{fig:Gt}).
%---------------------------------------------------------------------
%   Fig.5
%-------------------------------------------------------------------- 
%%%%%%%%%%%%%%%%%%%%%%%%%%%%%%%%%%%%%%%%%%%%%%%%%%%%%%%%
Therefore,  although  the spin gap still exists
for finite $U'$, we can expect,
by smooth extrapolation to $U'\rightarrow-\infty$, 
that the divergence in the above RG flows
may be suppressed, leading to the massless spin modes.

We should mention here that  the above treatment  
would fail at some negative critical value
$U'= U'_c$ if we only increase $|U'|$ by keeping
$U$ fixed.  Namely, for  $|U'| > U'_c$, the charge mode becomes 
massless, and then we cannot use the model as a weak-coupling  
analog of the depleted lattice.  This pathological result may  
come from the fact that 
the extremely strong attractive $U'$ may overwhelm 
the repulsive $U$ completely, resulting in  the spin bound states
similarly to the attractive Hubbard model and hence making the 
charge mode massless.
So, it may be necessary to take a strong-coupling limit by 
appropriately tuning the ratio of $U$
and $U'$ in the present model with attractive $U'$.

%-----------------------------------------------------------------
%                     III.C.
%----------------------------------------------------------------
\subsection{Discussions}

Within the above analysis for the effective weak coupling model, it 
turns out that the model is still massive for any finite coupling,
though the tendency to gapless excitations 
is seen as the singlet bound state becomes strong. 
This may suggest  that it is not straightforward to
incorporate the effects of depletion 
in a conventional continuum limit. 
In this sense, we should say that our analysis in this section is not  
satisfactory to describe  massless excitations
in the depleted spin model.  In this connection,
we would like to note here that the present RG analysis is 
based on the effective 
weak-coupling ``electron'' model because we believe that this
electron model belongs to the same universality
class of the original spin model. In fact
we have numerically solved the RG equations not only for 
the spin sector but also for the charge sector, and found that all the
trajectories flow to the strong coupling fixed points. 
Furthermore, as an extension, one could examine
whether the above situation is improved 
if we solve the RG equations only for spin sectors
by discarding the charge degrees of freedom,
which should be a more accurate approximation 
for the original spin model.
Unfortunately, even if we concentrate on the RG equations
for the spin sector, we still encounter the problem that 
the spin sector is  massive since the dimension of the inter-chain
couplings reduce to 1, as can be seen from Table \ref{t:Bos}. 
In this sense, a 
conventional weak-coupling bosonization approach, in which 
two chains are assumed to be weakly coupled with each other,
does probably not allow a straightforward description of the low-energy
properties of the depleted model.
%%Namely, considering the fact that there still exist the Umklapp 
%%process in eq.(\ref{HubInt}), which originates from the additional Hubbard
%%interaction $U'$, we learn that the charge sector form gaps and 
%%we can make them frozen. 
%%Then, from Table\ref{t:Bos}, we see that the scaling dimension of 
%%the inter-chain couplings changes from 2 to 1.
%%Therefore, spin sector has also gaps, proportional to $\widetilde J$.
%%These features are essentially the same as those of usual 
%%non-depleted spin ladders.
%%%%%%%%%%%%%%%%%%%
%%Therefore these results imply that our weak-coupling approach
%%in this section, which is based on a conventional scheme for coupled 
%%chains, is not satisfying in that the effects of depletion 
%%are not sufficiently incorporated in the results.
We thus have to find or develop a more effective way
to incorporate the effect of depletion correctly.
This problem will be  discussed in the following section.

%%Although from our one-loop RG calculation presented above,
%%it is not easy to draw a definite conclusion for
%%the strong coupling regime, all the results 
%%obtained both for attractive and repulsive
%%$U'$ suggest that in the limit $|U'|\rightarrow
%%\infty$, where the model may be reduced to the depleted 
%%spin model, the spin sector approaches 
%%the fixed point of massless spin excitations. 

%-----------------------------------------------------------------
%                     IV
%----------------------------------------------------------------
\section{Non-linear sigma model approach}

%%In the previous section we have seen the tendency towards
%%the formation of gapless excitations in the related 
%%Hubbard type model, although it still has a spin gap
%%in the region of interaction strength we have considered.
In the previous section, we have seen 
that  it is not straightforward to incorporate
the effect of depletion in an ordinary weak-coupling
approach. 
%%%in which two chains are weakly coupled each other,
%%%although the tendency towards
%%%the formation of gapless excitations in the related 
%%Hubbard type model is observed.
In particular,  it is quite difficult in this weak-coupling model
 to figure out which kind of spin 
excitations would actually become massless.
%% when  the effect of the depletion is
%%%completely taken into account.  
To clarify this point, and also to confirm 
that the massless modes are actually realized, we would like to
consider a different field theoretical approach  
%%now to consider the low-energy excitations further 
%%by a spin wave analysis and 
by using non-linear sigma model 
techniques.\cite{Hal,Rev,NlsLad,NAGAOSA}

Let us reconsider here the essential ingredients important for 
low energy excitations in our model.
When applying the LSM theorem in sec.II, we have seen that
the presence of unpaired spins associated with
vacant sites play a crucial role. All other spins are essentially
bound into local singlet pairs and cannot contribute to the very
low energy spectrum. The vacant sites break the uniformity of the
system by introducing these unpaired spin degrees of freedom
and by the local polarization of a staggered moment in their vicinity.
A short-ranged effective 
interaction among the unpaired spins is mediated through
polarization of the remaining spins, which is (anti-)ferromagnetic 
for $m=2~(4)$ mod 4. These interactions are weak and are assumed
to introduce a gapless spectrum. In a weak coupling approach
in the previous section, we have started 
with ordinary boson fields by simply applying the 
conventional bosonization schemes to each chain,
and have not explicitly taken into account the above properties
 in the beginning. 
Therefore, in order to describe the massless mode 
better, we need 
a rather complicated combination of these boson fields
by incorporating higher-order interactions.
This may be a reason why it was difficult in the 
weak coupling model to explicitly construct the massless spin 
excitations. 

Based on these observations,  
we would like to consider below a complementary approach
which allows us to describe the formation of massless 
excitations more easily. We will concentrate on the 
regime of comparatively small $m$ where the
background of the staggered magnetization plays an important role.
We first perform a spin-wave analysis of the depleted model
to clarify the nature of spin excitations.  We then convert 
the lowest spin mode to the non-linear sigma model, finding that
the massless excitations are indeed
generated for the depleted model.

%-----------------------------------------------------------------
%                     IV.A.
%-----------------------------------------------------------------
\subsection{Spin-Wave Analysis}

\begin{figure}[h]
\epsfxsize=7cm %%% 8 is suitable for 2 column
\centerline{\epsfbox{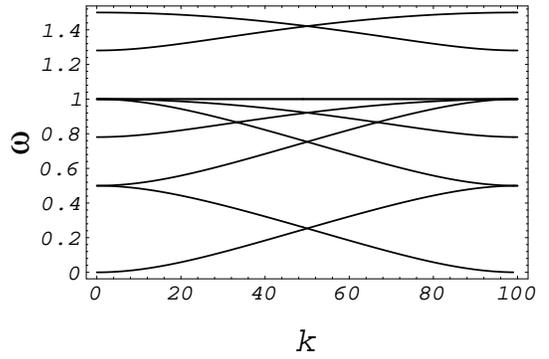}} 
\vspace{0.5cm}
\caption{Dispersion of $m=6$ system with $J=1$
as functions of $k$ with $N=600$.
In the usual notation, $k=100$ point corresponds to the momentum
$2\pi i\times100/600=\pi/3$. See eq.(\ref{DefGam})}
\label{fig:SpiWav6}%--------------------------------------------------
\end{figure}

\begin{figure}[h]
\epsfxsize=7cm %%% 8 is suitable for 2 column
\centerline{\epsfbox{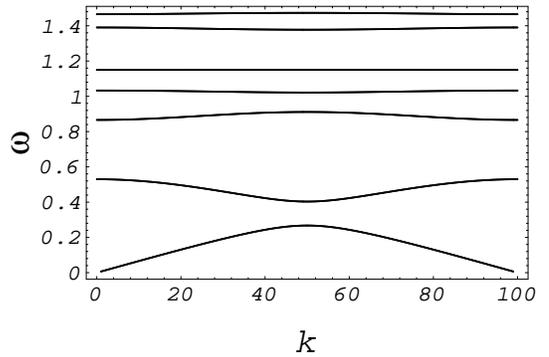}} 
\vspace{0.5cm}
\caption{Dispersion of $m=8$ system with $j=1$
as functions of $k$ with $N=800$.
In the usual notation, $k=100$ point corresponds to the momentum
$2\pi i\times100/800=\pi/4$.}
\label{fig:SpiWav8}%--------------------------------------------------
\end{figure}

We consider now a regularly 
depleted spin ladder with rather small
$m$. Then the polarization cloud of staggered magnetization around
each vacancy have a strong overlap with each other. Assuming coherent,
though slightly inhomogeneous staggered magnetization throughout
the whole ladder we may use
the Holstein-Primakoff mapping. Because after regular 
depletion (period $m$) the unit cell of the ladder is large we
represent the spin operators by $2m$ kind of bosons $a_l^{(i)}$ 
with $i=1,2$ and $l=0,1,\cdots m-1$ 
(We here include the depleted sites $i=1,~l=0$ and $i=2,~l=m/2$ 
for simplicity. But their sector easily decouples, which has always
zero-energy),
%-----------------------------------------------------------------
%   Holstein-Primakoff
%------------------------------------------------------------------
\begin{eqnarray}
&&S^{(i)z}_{mj+l}=
\pm\left(\frac{1}{2}-n_l^{(i)}(mj+l)\right),\\
&&S^{(i)+}_{mj+l}=\left\{
\begin{array}{c}
\sqrt{1-n_l^{(i)}(mj+l)}a_l^{(i)}(mj+l)\\
a_l^{(i)\dagger}(mj+l)\sqrt{1-n_l^{(i)}(mj+l)}\end{array}\right. \quad 
\end{eqnarray}
where $n_l$ is the number operator. In (4.2)
we take the upper (lower) relation for $i+l=$ odd (even).
In the momentum representation,
%------------------------------------------------------------------
%   Fourier trans. of boson op.
%---------------------------------------------------------------
\begin{equation}
a_l^{(i)}(mj+l)=\frac{1}{\sqrt{M}}\sum_{k=0}^{M-1}
\exp\left[\frac{2\pi i}{N}(mj+l)k\right]a_l^{(i)}(k),
\end{equation}
the Hamiltonian becomes, 
up to quadratic order of the boson operators,
%-------------------------------------------------------------------
%   spin wave hamiltonian
%---------------------------------------------------------------------
\begin{equation}
H=\sum_k\sum_{i,j=1}^{2m}\left(
a_i^\dagger h_{ij}a_j+a_ih_{ij}a_j^\dagger+
a_i^\dagger \Delta_{ij}a_j^\dagger+a_i\bar{\Delta}_{ij}a_j\right) 
\label{SpiWavHam}%---------------------------------------------------
\end{equation}
where we have omitted some trivial constant terms. 
We have used a simplified notation 
(See Fig.\ref{fig:SpiLad} for this numbering),
$a_0^{(1)}(k)=a_1$, $\cdots,a_{m-1}^{(1)}(-k)=a_{m}$,
$a_0^{(2)}(-k)=a_{m+1}$, $\cdots$, $\cdots$,
$a_{m-1}^{(2)}(k)=a_{2m}$, 
and the $2m\times 2m$ matrices $h$ and $\Delta$ are given in 
Appendix D.
By the Bogoliubov transformation
%-------------------------------------------------------------------
%   Bogoliubov trans.
%--------------------------------------------------------------------
\begin{equation}
a_i=\sum_{j=1}^{2m}\left(
U_{ij}b_j+\bar{V}_{ij}b_j^\dagger
\right),
\end{equation}
which should satisfy the relations
%--------------------------------------------------------------------
%   constraints
%-------------------------------------------------------------------
\begin{equation}
UV^\dagger-\bar{V}U=0,\quad UU^\dagger-\bar{V}V^T={\boldmath 1},
\end{equation}
we can reach the diagonal form of the Hamiltonian,
up to constant terms,
%--------------------------------------------------------------------
%   diagonalized spin wave hamiltonian
%---------------------------------------------------------------------
\begin{equation}
H=\sum_{j=1}^{2m}\omega_jb_j^\dagger b_j + \hbox{const.} ,
\end{equation}
In Figs.\ref{fig:SpiWav6},\ref{fig:SpiWav8}, 
we show the dispersion $\omega_i$ as a function of $k$,
in which we omitted the zero-energy modes 
associated with depleted spins.
We can observe that several bands appear because of the periodic
depletion. 
It is remarkable that within this approach the difference
between the cases $m=2$ and $m=4$ (mod 4) is reproduced clearly.
%---------------------------------------------------------------------
%   Fig.7
%-------------------------------------------------------------------- 
%---------------------------------------------------------------------
%   Fig.8
%-------------------------------------------------------------------- 
As discussed in section II, the ground state of the former case is
ferromagnetic (partially spin polarized), which is reflected in the 
quadratic dispersion of the lowest band for the small momenta 
(Fig.\ref{fig:SpiWav6}).
In contrast the latter behaves quite differently, displaying
a linear low-energy spectrum.
Therefore, the lowest band is indeed due to the interaction between
unpaired spins, as discussed at the beginning of this section.
%%%%%%%%%%%%%%%%%%%%%%%%%%%%%%%%%%%%%%%%%%%%%%%%%%%%%%
Though the lowest spin mode does not have an excitation gap 
in both cases, one cannot naively conclude a gapless spectrum for the
system, because higher-order quantum corrections play an important
role to determine whether or not the system is indeed gapless. 
This problem is investigated in the subsection B. 
%%%%%%%%%%%%%%%%%%%%%%%%%%%%%%%%%%%%%%%%%%%%%%%%%%%%%%%%%%%%%%%%%

We would like to mention here that although
our approach is valid for the case of small $m$, the qualitative
properties of the low-energy spin excitation spectrum are the same
for large $m$. In this case, however, a quantitative description would
have to include the effects of the coupling among elementary 
excitations beyond
(4.4) and (4.7). It is not our aim to address this problem here.

%-------------------------------------------------------------------
%                         IV.B.
%------------------------------------------------------------------
\subsection{Mapping to the Non-Linear Sigma Model}

Based on the above spin-wave analysis, we would like to now consider
the lowest band due to the unpaired spins and to discuss the
role of quantum effects by employing sigma model techniques.
%%%?Still we could completely rule out the possibility that there is
%%%a spin gap ?????. 
The results in the previous subsection suggest that 
the system with $m=4$ (mod 4) can be mapped to the non-linear sigma model, 
as is usually the case for uniform spin chains as well as 
uniform spin-ladder systems.
We restrict ourselves to such cases in this section.

We use here the coherent state path-integral method.\cite{Coh}
The partition function $Z={\rm tr}\exp(-\beta H)$ can be represented by
$Z=\int\prod_{i,j}{\cal D}[\mu(\bN^{(i)}(j))]\exp(-S)$,
where the unit vector $\bN^{(i)}(j)$ specifies the coherent state of 
the spin at the $j$th site of the $i$th chain such that
$\langle\bN^{(i)}(j)|\bS_j^{(i)}|\bN^{(i)}(j)\rangle=s\bN^{(i)}(j)$
with $s=1/2$, and $\mu(\bN)$ is the invariant measure
$\int d\mu(\bN)|\bN\rangle\langle\bN |=1$.
The Euclidean action $S$ is 
%--------------------------------------------------------------------
%   total action
%------------------------------------------------------------------
\begin{equation}
S=S_B+S_{\|}+S_{\bot},
\end{equation}
where
%--------------------------------------------------------------------
%   each action
%-------------------------------------------------------------------
\begin{eqnarray}
&&S_B=-is\sum_{i=1}^2\sum_{j=0}^{N-1}(1-\theta_j^{(i)})
\omega [\bN^{(i)}(j)],
\nonumber\\
&&S_\|=s^2\int_0^\beta\!\! d\tau\sum_{i=1}^2\sum_{j=0}^{N-1}
(1-\theta_j^{(i)})(1-\theta_{j+1}^{(i)})
\nonumber\\
&&\qquad\qquad\qquad\qquad\qquad
\times\bN^{(i)}(j)\cdot\bN^{(i)}(j+1),
\nonumber\\
&&S_\bot=Js^2\int_0^\beta\!\! d\tau\sum_{j=0}^{N-1}
(1-\theta_j^{(1)})(1-\theta_{j}^{(2)})
\nonumber\\
&&\qquad\qquad\qquad\qquad\qquad
\times\bN^{(1)}(j)\cdot\bN^{(2)}(j) .
\end{eqnarray}
Here $\omega [\bN]$ is the Berry phase defined by
%--------------------------------------------------------------------
%   Berry phase
%-------------------------------------------------------------------
\begin{equation}
\omega [\bN]=\int_0^\beta\!\! d\tau\!\!\int_0^1\!\! du
\bN\cdot(\partial_\tau\bN\times\partial_u\bN) .
\end{equation}
After staggering the configuration
%-------------------------------------------------------------------
%   staggering
%------------------------------------------------------------------
\begin{equation}
\bN^{(i)}(j)=(-)^{i+j+1}\bn^{(i)}(j),
\end{equation}
we suppose
%---------------------------------------------------------------------
%   m and l fields
%-------------------------------------------------------------------
\begin{equation}
\bn^{(i)}(j)=\bm(j)+(-)^{i+j+1}a_0\bl(j).
\end{equation}
In what follows, we assume the fields $\bm$ and $\bl$ are 
sufficiently smooth functions.
If we want to describe 
not only the lowest band but also the massive bands,
we must include several kinds of fields.
However, we concentrate on the lowest one only.
We expect that the above (non-trivial) assumption are sufficient for this
description. 

Let us now derive the continuum limit of the action.
First we 
consider the Berry phase term and divide it into two contributions,
%-----------------------------------------------------------------
%   S_B
%-----------------------------------------------------------------
\begin{eqnarray}
S_B&=&-is\sum_{i=1}^2\sum_{j=0}^{N-1}
(-)^{i+j+1}\omega [\bn^{(i)}(j)]
\label{Ber1}\\ %------------------------------------------------
&&+is\sum_{i=1}^2\sum_{j=0}^{N-1}(-)^{i+j+1}
\theta_j^{(i)}\omega [\bn^{(i)}(j)] .
\label{Ber2}%---------------------------------------------------
\end{eqnarray}
Taking now the continuum limit in each chain in eq.(\ref{Ber1}),
the topological terms cancel out
due to the factor $(-)^i$, and we find
%------------------------------------------------------------------
%   S_B uniform
%---------------------------------------------------------------
\begin{equation}
(\ref{Ber1})=2is\int\!\! d^2x\bl\cdot(\bm\times\partial_2\bm),
\end{equation}
where $x_1=x~(=a_0j)$ and $x_2=\tau$.
Then the contribution from eq.(\ref{Ber2}) is evaluated 
%-----------------------------------------------------------------
%   S_B  unpaired 1
%----------------------------------------------------------------
\begin{eqnarray}
(\ref{Ber2})&=&is\sum_{j=1}^{M-1}\left\{
\omega[\bn^{(1)}(mj)]-\omega[\bn^{(2)}(mj+m/2)]\right\}
\nonumber\\
&\sim&is\sum_{j=1}^{M-1}\int\!\! d\tau
(-)\delta\bn^{(1)}(mj)
\nonumber\\
&&\qquad\qquad\cdot
\left[\bn^{(1)}(mj)\times\partial_\tau\bn^{(1)}(mj)\right],
\end{eqnarray}
where
%--------------------------------------------------------------------
%   replace the difference to differential
%-------------------------------------------------------------------- 
$\delta\bn^{(1)}(mj)\equiv\bn^{(1)}(mj)-\bn^{(2)}(mj+\frac{m}{2})
\sim-\frac{m}{2}a_0\partial_x\bm(mj)
+2a_0\bl(mj)+O(a_0^2)$. 
%%%%%%%%%%%%%%%%%%%%%%%%%%%%%%%%%%%%
We have assumed that the short-range AF order is strong
enough to justify our continuum limit.
%%%%%%%%%%%%%%%%%%%%%%%%%%
As noted above in this subsection, we here concentrate on
the case $m=4$ (mod 4), because the above equation is valid 
only for $m/2=$ even.
%%%%%%%%%%%%%%%%%%%%%%%%%%%%%%%%%%%%%
Substituting this relation, we have
%-------------------------------------------------------------------
%   S_B unpaired 2
%-------------------------------------------------------------------
\begin{eqnarray}
(\ref{Ber2})=&&
-\frac{is}{2}\int\!\! d^2x\bm\cdot(\partial_1\bm\times\partial_2\bm)
\nonumber\\
&&-\frac{2is}{m}\int\!\! d^2x\bl\cdot(\bm\times\partial_2\bm).
\end{eqnarray}

Next, we evaluate the intra-chain interaction.
For example, the contribution from chain 1 is
%----------------------------------------------------------------
%   S_\|^1
%----------------------------------------------------------------
\begin{eqnarray}
S_\|^{(1)}&=&\frac{s^2}{2}\int\!\! d\tau
\sum_{j=0}^{M-1}\sum_{l=1}^{m-2}
\left[\bn^{(1)}(mj+l)-\bn^{(1)}(mj+l+1)\right]^2
\nonumber\\
&\sim&\frac{a_0s^2}{2}\sum_{l=1}^{m-2}\int\!\!\frac{dx}{m}
\int\!\! d\tau
\left[(\partial_x\bm)^2-4(-)^l\partial_x\bm\cdot\bl+4\bl^2\right]
\nonumber\\
&=&\frac{a_0s^2}{2}\frac{m-2}{m}\int\!\! d^2x
\left[(\partial_x\bm)^2+4\bl^2\right] .
\end{eqnarray}
%%%%%%%%%%%%%%%%%%%%%%%%%%%%%%%%%%%%%%%%%%%%%%%%%%%%
This type of approximation may be justified for rather 
small $m$ where we can assume that
the region between the vacant sites is clearly dominated by the
staggered spin polarization. Then all the intermediate spins
vary very slowly and essentially in phase with each other.
The contribution from chain 2 is exactly the same, 
and we find
%------------------------------------------------------------------
%   S_\|
%---------------------------------------------------------------------
\begin{equation}
S_\|=as_0^2\left(1-\frac{2}{m}\right)\int\!\! d^2x
\left[(\partial_x\bm)^2+4\bl^2\right].
\end{equation}
Inter-chain interaction is similarly evaluated 
%-------------------------------------------------------------------
%   S_\bot
%------------------------------------------------------------------
\begin{equation}
S_\bot=2Ja_0s^2\left(1-\frac{2}{m}\right)\int\!\! d^2x\bl^2 .
\end{equation}

Collecting terms thus obtained and integrating the field $\bl$ out, 
we have $S=\int d^2x{\cal L}$, where
%----------------------------------------------------------------
%   final laglangian
%----------------------------------------------------------------
\begin{eqnarray}
{\cal L}=&&\frac{1}{2g}\left[
v(\partial_1\bm)^2+\frac{1}{v}(\partial_2\bm)^2\right]
\nonumber\\
&&\qquad-\frac{\theta}{8\pi}
\epsilon_{\mu\nu}\bm\cdot(\partial_\mu\bm\times\partial_\nu\bm) ,
\end{eqnarray}
where
%-----------------------------------------------------------------
%   \theta,g,v
%------------------------------------------------------------------
\begin{eqnarray}
&&\theta=2\pi is=\pi i ,\nonumber\\
&&g=\frac{m}{s(m-1)}\sqrt{1+r/2} ,\nonumber\\
&&v=2Ja_0s\frac{m-2}{m-1}\sqrt{1+r/2} ,
\label{Tgv}%-----------------------------------------------------
\end{eqnarray}
with
$r=1/J~(=J_\|/J_\bot)$.

It should be noted here  that the topological term 
in the above action appears with the 
coefficient $\pi i$, which suggests that the present system
should be massless, as we expected.
Therefore, combining this result with 
the analyses in secs. II and III,
we end up with the conclusion that the present 
depleted system with spin singlet ground state 
should have massless spin excitations,
belonging to the same universality class as 
the uniform $s=1/2$ Heisenberg chain.

Before concluding this section, a brief 
comment on the case of larger $m$ is in order. 
As we mentioned, our present approach is valid for the case of
rather small $m$.  
If we naively set $m\rightarrow\infty$, 
bulk quantities such as $g$ and $v$ in eq.(\ref{Tgv})
reduce to the formula 
derived by Dell'Aringa et. al.\cite{NlsLad}
for the uniform non-depleted spin ladders,
while the coefficient $\theta$ remains $\theta=\pi i$.
%%%%%%%%%%%%%%%%%%%%%%%%%%%%%%%%%%%%%%%%%%%%%%%%%%
This result seems somewhat paradoxical at first sight. However,
one should note that the continuum limit in our approach means 
$M\equiv N/m\rightarrow\infty$.  Therefore, however large $m$ may 
be, the system always has the periodic array of impurities with
finite density (infinite number of impurities in the thermodynamic 
limit), which may form the massless mode inside of the Haldane 
gap even for the limit of large $m$. This massless mode 
should be smoothly extrapolated from the massless mode obtained for
the  small $m$ case. In this sense, the limit $m\rightarrow\infty$ 
contains a subtle problem. We think that in the present
approximation the topological term $\theta=\pi i$ correctly 
reflects this
massless mode whereas the values of $g$ and $v$ are determined
by the host part with the Haldane gap in the large $m$ limit.
Therefore, to apply our analysis to the low-energy physics in 
the large $m$ case consistently, we need to improve systematically 
the evaluation of the bulk quantities of $g$ and $v$.
In fact, we have checked in a preliminary calculation that
the estimated values of the above bulk quantities are improved if
we include not only the lowest spin mode but also higher modes.  From 
these discussions, we believe that the present conclusion
drawn from the above treatment applies also to the case of 
larger $m$  at least qualitatively, although the approximation 
becomes worse when $m$ becomes large.  To improve our
results in the large $m$ case is an open problem to be explored
in the future study. 
%%%%%%%%%%%%%%%%%%%%
Note that in the case of larger $m$, an appropriate scheme to treat  
low-energy excitations 
has been considered recently 
by Nagaosa et al. in the context of a randomly depleted spin ladder
\cite{NAGAOSA}. 

%-----------------------------------------------------------------
%                     V
%-----------------------------------------------------------------
\section{Summary}

We have considered a class of the models for 
regularly depleted two-leg spin ladder systems,
and investigated their low-energy properties. 
It has been proved rigorously 
that by the depletion
this special class of the models generates 
a new state which is characterized either by the presence of
gapless excitations or a degenerate ground state.
To see whether the massless states are really produced
by the depletion, we have investigated low-energy properties
in terms of two different field theoretical approaches.

We have first proposed a scheme to describe low-energy 
properties by applying renormalization 
group methods to an effective weak-coupling
model of Hubbard type, in which 
the effect of  the depletion is incorporated
in the strong coupling limit of the on-site interaction. 
This model has been analyzed in a weak coupling approximation
based on the one-loop RG method.
%%%%%%%%%%%%%%%%%%%%%%%%%%%%%%%%%%%
Although we have seen how the tendency to a massless state 
is developed when the effect of the depletion becomes strong,
it has been found that our model is  still massive for any finite 
couplings. This implies that it is not easy to 
describe the effect of depletion 
in a naive continuum limit by a conventional method.
%%%%%%%%%%%%%%%%%%%%%%%%%%%%%%%%%%%%%%%%%%%%%%%

To clarify this point,
we studied the depleted spin model by mapping it to the
non-linear sigma model. 
We have found via  the spin-wave analysis
that the dispersion relation of the lowest  
spin wave mode has indeed a linear spectrum in the case 
of the singlet ground state, which allowed us to
mapping to the non-linear sigma model with a topological term.
It has been shown that the coefficient of this topological term is
$\pi i$, which coincides with that of the models which have massless 
spin excitations such as the spin 1/2 Heisenberg chain. 
By combining all the analyses in secs.II, III and IV,
we thus conclude that 
the periodic depletion of the present spin ladder systems
should produce massless spin excitations.

Finally we note that the drastic change 
from the spin-gap state to the 
gapless state observed in the present model
reflects that the phase coherence 
in the wave function is quite sensitive to the depletion.  
In this sense, the formation of the gapless spin state 
in our periodic model have the same basic origin as that 
with random  non-magnetic impurities, although the low-temperature
thermodynamics of the two is rather different. \cite{FURUSAKI} 

\acknowledgements        

The authors would like to thank A. Furusaki for helpful
discussions. M.S. is grateful to the Swiss Nationalfonds for
financial support (PROFIL-Fellowship). 
This work is partly supported by the Grant-in-Aid from the Ministry of
Education, Science and Culture, Japan.

\appendix
%---------------------------------------------------------------------
%                         APPENDIX A.
%---------------------------------------------------------------------
\section{Continuum limit of interactions}

In this Appendix, we derive the continuum limit of interactions
(\ref{HubInt}) and (\ref{InterCha}).
First, the continuum limit of the intra-chain interaction is 
calculated as 
%-----------------------------------------------------------------
%   intra-chain int.
%------------------------------------------------------------------
\begin{eqnarray}
H_{\rm int}^{(i)}=\int dx&&\left(Ua+U'a\theta_j^{(i)}\right)
\bigl(2J_{iL}^0J_{iR}^0-2J_{iL}^aJ_{iR}^a
\nonumber\\
&&+e^{4ik_Fx}\psi_{iL\uparrow}^\dagger\psi_{iR\uparrow}
\psi_{iL\downarrow}^\dagger\psi_{iR\downarrow}+{\rm h.c.}\bigr)
\end{eqnarray}
Note here that $\theta_j$ defined by eq.(\ref{TheJ}) can be rewritten as
%-------------------------------------------------------------------
%   \theta_j
%--------------------------------------------------------------------
\begin{equation}
\theta_j=\frac{1}{m}\sum_{l=1}^{m}e^{2\pi ilj/m}.
\end{equation}
Therefore, various kinds of terms behave as
%---------------------------------------------------------------------
%   oscillation
%-----------------------------------------------------------------------
\begin{eqnarray}
&&e^{4ik_Fx}=e^{2\pi i(1+1/m)j}
=\hbox{(ossillating terms)},
\nonumber\\
&&\theta_j^{(i)}=\frac{1}{m}+\hbox{(ossillating terms)},
\nonumber\\
&&\theta_j^{(i)}e^{4ik_Fx}=(-)^{i-1}\frac{1}{m}
 +\hbox{(ossillating terms)},
\label{adddd}
\end{eqnarray}
where $i$ in $\theta^{(i)}$ is $i=1,2$.
Consequently, if we neglect oscillating terms, we have
eq.(\ref{HubInt}).

Next consider the inter-chain coupling.
By using the continuum limit of the spin operator,
%---------------------------------------------------------------------
%   spin operator
%----------------------------------------------------------------------
\begin{equation}
S_j^{(i)a}/a_0\rightarrow J_{iL}^a+J_{iR}^a+
\left(e^{2ik_Fx}N_i^a+\hbox{ h.c.}\right),
\end{equation}
where
\begin{equation}
N_i^a=\psi_{iL\alpha}^\dagger\frac{\sigma^a_{\alpha\beta}}{2}
\psi_{iR\beta}.
\end{equation}
the inter-chain coupling terms are calculated as
%---------------------------------------------------------------------
%   inter-chain
%---------------------------------------------------------------------
\begin{eqnarray}
H_{\rm coup}=&&\int dx
\left(1-\theta_j^{(1)}\right)\left(1-\theta_j^{(2)}\right)
\nonumber\\\times
&&\Bigl[J_{1L}^aJ_{2R}^a+J_{2L}^aJ_{1R}^a
+\left(N_1^aN_2^{a\dagger}+\hbox{h.c.}\right)
\nonumber\\
&& ~+(2k_F \hbox{ oscillating terms})
\nonumber\\
&& ~+(4k_F \hbox{ oscillating terms})\Bigr].
\end{eqnarray}
Note that
%--------------------------------------------------------------------
%   two \theta
%---------------------------------------------------------------------
\begin{equation}
\left(1-\theta_j^{(1)}\right)\left(1-\theta_j^{(2)}\right)
=1-\frac{2}{m}\sum_{l=1}^{m/2}e^{4\pi ilj/m}.
\end{equation}
Therefore, we have formulae similar to
(\ref{adddd}),
%--------------------------------------------------------------------
%   oscillation
%---------------------------------------------------------------------
\begin{eqnarray}
&&\left(1-\theta_j^{(1)}\right)\left(1-\theta_j^{(2)}\right)
=1-\frac{2}{m}+\hbox{ (oscillating terms),}
\nonumber\\
&&\left(1-\theta_j^{(1)}\right)\left(1-\theta_j^{(2)}\right)
e^{2ik_Fx}=\hbox{ (oscillating terms)},
\nonumber\\
&&\left(1-\theta_j^{(1)}\right)\left(1-\theta_j^{(2)}\right)
e^{4ik_Fx}=\hbox{ (oscillating terms)}.
\nonumber\\
\end{eqnarray}
By expressing $N_1^aN_2^{a\dagger}$ by $L$, 
\begin{equation}
N_1^aN_2^{a\dagger}+\hbox{h.c.}=
\frac{1}{2}(L_L^aL_R^{\dagger a}+\hbox{h.c.})
-\frac{3}{2}(L_L^0L_R^{\dagger 0}+\hbox{h.c.}),
\end{equation}
we end up with eq.(\ref{InterCha}).

%---------------------------------------------------------------------
%                       APPENDIX B. OPE
%---------------------------------------------------------------------
\section{Operator product expansion}
The basic operator product expansion (OPE) for Fermi fields is 
%---------------------------------------------------------------------
%   [1/2]*[1/2]
%--------------------------------------------------------------------
\begin{equation}
\psi_{iL\alpha}^\dagger(z)\psi_{jL\beta}(w)
\sim\frac{\delta_{ij}\delta_{\alpha\beta}}{2\pi(z-w)} ,
\end{equation}
where $z=v_F\tau+ix$.
Here and in what follows, we neglect regular terms.
Operator products should be normal-ordered,
though we do not explicitly indicate this. 
Similar formulae hold for right-moving currents, by replacing
$z\rightarrow\bar z$.
Define $f^{\mu\nu\lambda}$ and $d^{\mu\nu\lambda}$ as
$[\sigma^\mu/2,\sigma^\nu/2]=f^{\mu\nu\lambda}\sigma^\lambda/2$
and
$\sigma^\mu\sigma^\nu=d^{\mu\nu\lambda}\sigma^\lambda$.
Namely,
$f^{0\mu\nu}=f^{\mu0\nu}=f^{\mu\nu0}=0,~f^{abc}=i\epsilon^{abc}$
and
$d^{0\mu\nu}=d^{\mu0\nu}=d^{\mu\nu0}=\delta^{\mu\nu},
~d^{abc}=\delta^{ab}\delta^{c0}+i\epsilon^{abc}$.
Then we have
%-----------------------------------------------------------------------
%   [1]*[1]
%---------------------------------------------------------------------
\begin{eqnarray}
&&J_{iL}^\mu(z)J_{jL}^\nu(w)\sim
\delta_{ij}\left[\frac{\delta^{\mu\nu}}{8\pi^2(z-w)^2}
+\frac{f^{\mu\nu\lambda}J_{jL}^\lambda(w)}{2\pi(z-w)}\right],
\nonumber\\
&&L_{L}^\mu(z)L_{L}^{\nu\dagger}(w)\sim
\frac{\delta^{\mu\nu}}{8\pi^2(z-w)^2}
\nonumber\\
&&\qquad\qquad\qquad\qquad+\frac{d^{\mu\nu\lambda}J_{1L}^\lambda(w)
 -d^{\nu\mu\lambda}J_{2L}^\lambda(w)}{2\pi(z-w)},
\nonumber\\
&&L_{L}^{\mu\dagger}(z)L_{L}^{\nu}(w)\sim\quad(1\leftrightarrow2),
\nonumber\\
&&M_{iL}(z)M_{jL}(w)\sim\delta_{ij}
\left[
\frac{-1}{4\pi^2(z-w)^2}+\frac{-J_{jL}^0}{\pi(z-w)}
\right],
\nonumber\\
&&J_{iL}^\mu(z)L_L^\nu(w)\sim
\frac{(\delta_{i1}d^{\mu\nu\lambda}
 -\delta_{i2}d^{\nu\mu\lambda})L_L^\lambda(w)}{4\pi(z-w)},
\nonumber\\
&&J_{iL}^\mu(z)L_L^{\nu\dagger}(w)\sim\quad(1\leftrightarrow2),
\nonumber\\
&&J_{iL}^\mu(z)M_{jL}(w)\sim
\delta_{ij}\frac{-\delta^{\mu0}M_{jL}(w)}{2\pi(z-w)}.
\end{eqnarray}
The OPEs among various 4-Fermi interactions in the text follow from the 
above formulae.
Using these OPEs, we can derive the one-loop order RG equation as follows:
Suppose
%--------------------------------------------------------------
%   perturbed hamiltonian
%---------------------------------------------------------------
\begin{equation}
H=H_*-\sum_i\frac{g_i}{2\pi v_F}\int dxO_i ,
\end{equation}
where the operators $O_i$ represent the various 4-Fermi interactions 
with dimension 2 given above. 
The RG equations for such perturbations are given by
%---------------------------------------------------------------
%   RG equation
%-----------------------------------------------------------------
\begin{equation}
\dg{k}=(2-x_k)g_{k}+2\pi^2\sum_{i,j}C_{ijk}g_ig_j ,
\end{equation}
where $x_k$ is the dimension of the operator $O_k$ and 
$C_{ijk}$ is the OPE coefficients defined by
%---------------------------------------------------------------
%   OPE coefficient
%---------------------------------------------------------------
\begin{equation}
O_i(z)O_j(w)\sim\sum_k\frac{C_{ijk}}{|z-w|^2}O_k(w) .
\end{equation}

%---------------------------------------------------------------
%             APPENDIX C. BOSONIZATION
%------------------------------------------------------------------
\section{Bosonization}

The Fermi fields are bosonized as
\begin{eqnarray}
\psi_{iL\alpha}&=&
\frac{1}{\sqrt{2\pi a}}e^{-i\sqrt{4\pi}\varphi_{iL\alpha}},
\nonumber\\
\psi_{iR\alpha}&=&
\frac{1}{\sqrt{2\pi a}}e^{i\sqrt{4\pi}\varphi_{iR\alpha}}.
\end{eqnarray}
Define
\begin{eqnarray}
\varphi_{i\alpha}&=&\varphi_{iL\alpha}+\varphi_{iR\alpha},
\nonumber\\
\theta_{i\alpha}&=&\varphi_{iL\alpha}-\varphi_{iR\alpha},
\end{eqnarray}
and
\begin{eqnarray}
&&\varphi_{ic}=
\frac{1}{\sqrt{2}}(\varphi_{i\uparrow}+\varphi_{i\downarrow}),
\quad
\theta_{ic}=
\frac{1}{\sqrt{2}}(\theta_{i\uparrow}+\theta_{i\downarrow}),
\nonumber\\
&&\varphi_{is}=
\frac{1}{\sqrt{2}}(\varphi_{i\uparrow}-\varphi_{i\downarrow}),
\quad
\theta_{is}=
\frac{1}{\sqrt{2}}(\theta_{i\uparrow}-\theta_{i\downarrow}).
\end{eqnarray}
For ladder systems, it is convenient to define furthermore
the in- and out-of-phase Bose fields, 
\begin{eqnarray}
&&\varphi_c^\pm=
\frac{1}{\sqrt{2}}(\varphi_{1c}\pm\varphi_{2c}),
\quad
\theta_c^\pm=
\frac{1}{\sqrt{2}}(\theta_{1c}\pm\theta_{2c}),
\nonumber\\
&&\varphi_s^\pm=
\frac{1}{\sqrt{2}}(\varphi_{1s}\pm\varphi_{2s}),
\quad
\theta_s^\pm=
\frac{1}{\sqrt{2}}(\theta_{1s}\pm\theta_{2s}).
\end{eqnarray}
Then, except for the gradient terms, 
we have the bosonized form of the 
various interactions in the text,
summarized in the Table \ref{t:Bos}.

%-------------------------------------------------------------
%   Table of bosonization
%-------------------------------------------------------------
\begin{table}[h]
\begin{center}
\renewcommand{\arraystretch}{2}
\begin{tabular}{cc}
Coupling constants&Corresponding operators\\
\hline
$g_u$
 &$\frac{1}{\pi^2a^2}\sin\sqrt{4\pi}\varphi_c^+
  \sin\sqrt{4\pi}\varphi_c^-$\\
$g_\sigma$ 
 & $\frac{-1}{2\pi^2a^2}\cos\sqrt{4\pi}\varphi_s^+
  \cos\sqrt{4\pi}\varphi_s^-$\\
$g_{x\sigma}$ 
 & $\frac{1}{2\pi^2a^2}\cos\sqrt{4\pi}\varphi_s^+
  \cos\sqrt{4\pi}\theta_s^-$\\
$(g_{t\sigma}-g_{t\rho})/2$ 
 &$\frac{1}{2\pi^2a^2}\cos\sqrt{4\pi}\theta_s^-
  \cos\sqrt{4\pi}\varphi_c^-$\\
$-(g_{t\sigma}+g_{t\rho})$ 
 &$\frac{1}{2\pi^2a^2}\cos\sqrt{4\pi}\varphi_s^-
  \cos\sqrt{4\pi}\varphi_c^-$\\
$2g_{t\sigma}$ 
 &$\frac{-1}{4\pi^2a^2}\cos\sqrt{4\pi}\varphi_s^+
  \cos\sqrt{4\pi}\varphi_c^-$
\end{tabular}
\end{center}
\caption{Bosonized operators}
\label{t:Bos}%--------------------------------------------------
\end{table}

%---------------------------------------------------------------
%             APPENDIX C. SPIN WAVE HAMILTONIAN
%------------------------------------------------------------------
\section{Spin wave Hamiltonian}

Here we summarize the expressions for the matrices
$h$ and $\Delta$ in eq.(\ref{SpiWavHam}).
Let us restrict ourselves to the $J=1$ case.
First, the matrix $h$ is diagonal and given by
\begin{equation}
h_{ij}=\frac{1}{4}\delta_{ij}\times
\hbox{(\# of the nearest neighbour for $i$th site)},
\end{equation}
where $i$th site means its 
simplified notation used in eq.(\ref{SpiWavHam}).
For example, $m=6$ case, we have
\begin{equation}
h=\frac{1}{4}{\rm diag}(0,2,3,2,3,2,2,3,2,0,2,3),
\end{equation}
and $m=8$ case
\begin{equation}
h=\frac{1}{4}{\rm diag}(0,2,3,3,2,3,3,2,2,3,3,2,0,2,3,3).
\end{equation}

Interaction part $\Delta$ is constructed as follows:
First denote it by $m\times m$ submatrices
\begin{equation}
\Delta=\frac{1}{4}\left(
\begin{array}{cc}
\delta&1_m\\ 1_m&\bar{\delta}
\end{array}\right) ,
\end{equation}
where the matrix $\delta$ represents 
intra-chain coupling, defined by
\begin{equation}
\delta=\left(
\begin{array}{ccccccc}
0           &\gamma      &            &      &      &      &\bar{\gamma}\\
\gamma      & 0          &\bar{\gamma}&      &      &      &            \\
            &\bar{\gamma}& 0          &\gamma&      &      &            \\
            &            &\gamma      & 0    &      &      &            \\
            &            &            &      &\ddots&      &            \\
            &            &            &      &      & 0    &\gamma      \\
\bar{\gamma}&            &            &      &      &\gamma& 0     
\end{array}\right) ,
\end{equation}
with 
\begin{equation}
\gamma\equiv\exp(2\pi ik/N),
\label{DefGam}%----------------------------------------------------------
\end{equation}
and $1_m$ is the $m\times m$ unit matrix, corresponding to the
inter-chain coupling.
The matrix $\Delta$ thus defined corresponds to that of 
spin ladders without depletion. 
To include the effects of the depletion,
the non-zero elements in the first row and  column,
as well as in the $3m/2+1$th row and column should be set to 0. 
%-----------------------------------------------------------------------
%                             references
%----------------------------------------------------------------------


\begin{references}
\bibitem{Com1}
Because of this reason, we cannot reproduce the difference between
$m=4$ and $m=2$ mod 4 by the effective theory.
\bibitem{Com2}
Note that the other coupling constants  
$g_\sigma$  and $g_{x\sigma}$ for spin modes behave near the $l=l_c$
as
\begin{eqnarray}
&&g_\sigma\sim\frac{1}{2}(g_{t\sigma0}+g_{t\rho0})g_{t\sigma0}
\times\ln(l_c-l),
\nonumber\\
&&g_{x\sigma}\sim\frac{1}{2}(g_{t\sigma0}-g_{t\rho0})g_{t\sigma0}
\times\ln(l_c-l).
\nonumber
\end{eqnarray}
Therefore, divergent terms suppressed by $U'$
faster than $g_{t\sigma}$ and
$g_{t\rho}$ because of eq.(\ref{GT0}).
This is the reason why we could not see the divergent behavior in 
Fig.\ref{fg:flow2}.
\end{references}
\end{document}